\documentclass[sigconf, nonacm]{acmart}
\usepackage{caption}
\usepackage{multirow}
\usepackage{subcaption}
\usepackage{float}
\AtBeginDocument{%
  \providecommand\BibTeX{{%
    \normalfont B\kern-0.5em{\scshape i\kern-0.25em b}\kern-0.8em\TeX}}}

\setcopyright{acmcopyright}
\copyrightyear{2022}
\acmYear{2022}
\acmDOI{XXXXXXX.XXXXXXX}

\acmConference[Conference acronym 'XX]{Make sure to enter the correct
  conference title from your rights confirmation emai}{August 25--29,
  2024}{Barcelona, Spain}
\acmPrice{15.00}
\acmISBN{978-1-4503-XXXX-X/18/06}

\begin{document}


\title{Pfeed:
Generating near real-time personalized feeds using precomputed embedding similarities
}


\author{Binyam Gebre}
\affiliation{%
  \institution{Bol}
  \country{The Netherlands}}
\email{bgebre@bol.com}

\author{Karoliina Ranta}
\authornote{Both authors contributed to this work while working at Bol.}

\affiliation{%
  \institution{Booking.com}
  \country{The Netherlands}}
  \email{karoliina.ranta@booking.com}

\author{Stef van den Elzen}
\affiliation{%
  \institution{Eindhoven University of Technology}
  \country{The Netherlands}}
  \email{s.j.v.d.elzen@tue.nl}

\author{Ernst Kuiper}
\affiliation{%
  \institution{Bol}
  \country{The Netherlands}}
  \email{ekuiper@bol.com}

\author{Thijs Baars}
\authornotemark[1]
\affiliation{%
  \institution{Last Mile Solutions}
  \country{The Netherlands}}
  \email{thijs.baars@lastmilesolutions.com}
  
\author{Tom Heskes}
\affiliation{
  \institution{Radboud University Nijmegen}
  \country{The Netherlands}}
  \email{tom.heskes@ru.nl}

\renewcommand{\shortauthors}{Gebre, Ranta, van den Elzen, Kuiper, Baars, and Heskes}

\begin{abstract}
In personalized recommender systems, embeddings are often used to encode customer actions and items, and retrieval is then performed in the embedding space using approximate nearest neighbor search. However, this approach can lead to two challenges: 1) user embeddings can restrict the diversity of interests captured and 2) the need to keep them up-to-date requires an expensive, real-time infrastructure. In this paper, we propose a method that overcomes these challenges in a practical, industrial setting. The method dynamically updates customer profiles and composes a feed every two minutes, employing precomputed embeddings and their respective similarities. We tested and deployed this method to personalize promotional items at Bol, one of the largest e-commerce platforms of the Netherlands and Belgium. The method enhanced customer engagement and experience, leading to a significant 4.9\% uplift in conversions.

\end{abstract}


%
\maketitle

\section{Introduction}
Bol, like many other e-commerce platforms, faces the challenge of providing customers with an easy and efficient way to navigate their vast catalog and find products that match their customers' interests. The traditional approach of relying on customer controlled text-based search engines or browsing through categories is often limited and cumbersome, particularly during the customer's discovery phase. To overcome these limitations and enhance customers' overall discovery experience, Bol has launched personalized feeds called \textit{Top deals for you}, \textit{Top picks for you}, and \textit{New for you}. 

These personalized feed systems utilize a combination of the customer's historical and recent behavior to display the best recommendations on the customer's home page across both app and desktop platforms. In this paper, we present the methodology behind these feeds. We begin by presenting the challenges inherent to creating personalized feed systems. Subsequently, we delve into the prevailing industry approach (related work) that tackles these challenges, concluding with the presentation of our proposed solution and the evaluation outcomes.

\subsection{Four Challenges in Personalized Feed Systems}
Personalized feed systems can be viewed as search engines, where customers are the search queries and items in the catalog are the search results. In this view, there are four challenges that need to be overcome to provide customers with a personalized set of items that align with their interests and preferences: customer, item, candidate retrieval and ranking challenges.

\subsubsection{Customer representation challenge} 
Customers show complex behaviors while shopping on e-commerce sites before making a purchase, e.g.,  searching for items, viewing items, reading reviews, and making item comparisons. The challenge is distilling these interactions into a concise customer representation. In addition to their dynamic interactions, the representation may also need to incorporate  static attributes of customers, such as customer ID, gender, and clothing-size.

\subsubsection{Item representation challenge} 
Items have rich structured information such as item ID, title, description, specifications, and other metadata. Items also have historical customer interactions: views, clicks, customer ratings, reviews, etc. The item representation challenge is identifying the most relevant data for representing various items, a task complicated by two factors. The first is the diversity of item attributes. For instance, \textit{author} and \textit{title} are key attributes for books, whereas \textit{size} and \textit{gender} are more critical for fashion items. The second factor is the cold-start problem associated with new products; these have no historical interactions.

\subsubsection{Candidate retrieval challenge} 
Candidate retrieval entails determining which items best match a given customer's preferences. Here, the challenges are of two varieties: 1) training customer and item representations in the same embedding space and 2) the inference challenge, which aims to efficiently retrieve the best matches from a corpus containing millions to billions of items. 

\subsubsection{Ranking challenge} 
The candidate retrieval stage is followed by a ranking stage, where the retrieved candidates are re-ranked using a more complex model and more complex features of both the retrieved candidates and queries. The goal of this stage is to select and rank the top K items per customer (for example, the top 100 items) using learning-to-rank algorithms \cite{10.1145/2988450.2988454, 10.1145/3219819.3219823,10.5555/3172077.3172127, 10.1145/3442381.3450078}. 

In this paper, we focus on addressing the first three challenges: the \textit{customer representation}, \textit{item representation}, and \textit{candidate retrieval} challenges. 

\subsection{Our Contributions}
\begin{figure}[h]
  \centering
  \includegraphics[width=0.7\linewidth]{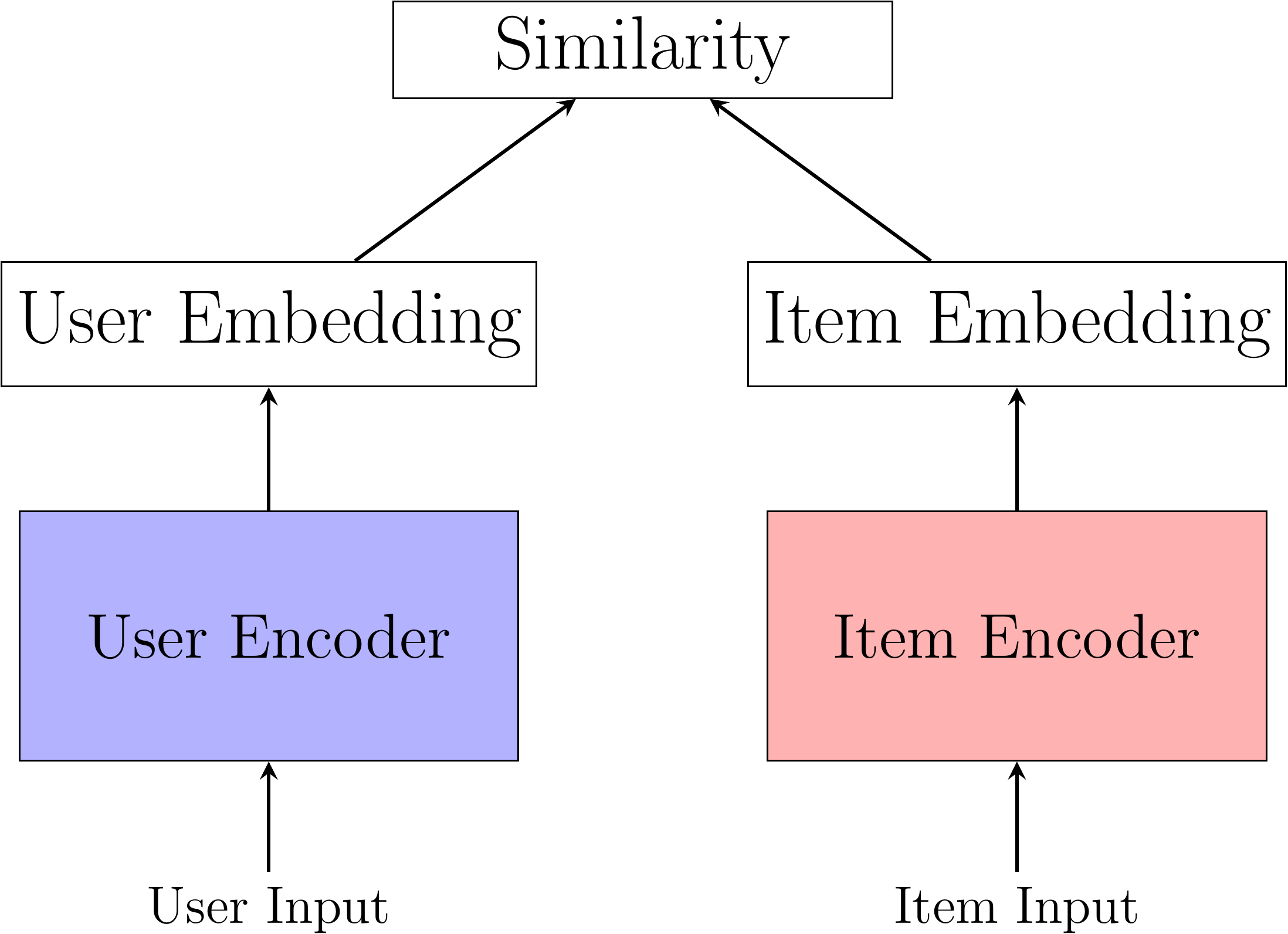}
  \caption{User-to-item framework: Single vectors from the user encoder limit representation and interpretability. Keeping them fresh demands high-maintenance infrastructure.}
  \label{fig:user-to-item-framework}
\end{figure}

The most dominant approach to tackling the aforementioned challenges relies on a user-item framework (see figure \ref{fig:user-to-item-framework}). Two neural networks, called dual encoders, are each trained to generate embeddings for user and item data \cite{10.1145/3534678.3539156, 10.1145/3394486.3403280,covington2016deep, 10.1145/3298689.3346996,wang2021personalized}. 
The user embedding model receives input in the form of a sequence or bag of interactions on items, along with context and user data \cite{wang2021personalized}. On the other hand, the item embedding model utilizes various item metadata types including item IDs \cite{covington2016deep, 10.1145/3534678.3539156, 10.1145/3394486.3403344} or output embeddings from pre-trained models \cite{10.1145/3219819.3219890,10.1145/3534678.3539156,10.1145/3394486.3403280}. 

However, despite its widespread use, the user encoding model in this framework has two significant drawbacks: the \textit{single vector representation bottleneck} and the \textit{high infrastructure and maintenance costs}.

\subsubsection{Single vector representation bottleneck}

Using a single vector to represent users introduces challenges due to the diversity and complexity of their interests, compromising both the capacity to accurately represent users and the interpretability of the representation by obscuring which interests are represented and which are not. While attempts to use multiple embeddings have been made to overcome these limitations, the exact number of vectors needed and the method for obtaining them remain topics of research \cite{10.1145/3357384.3357814,10.1145/3394486.3403280}.

\subsubsection{High infrastructure and maintenance costs}
Generating and maintaining up-to-date user embeddings requires substantial investment in terms of infrastructure and maintenance (see, for example, the SOAP platform from Meta \cite{zhang2023scaling}). Each new user action necessitates executing the user encoder to generate fresh embeddings and recommendations. Furthermore, the user encoder must be large in order to effectively model a sequence of interactions, leading to expensive training and inference requirements.

\begin{figure}[h]
  \centering
  \includegraphics[width=0.7\linewidth]{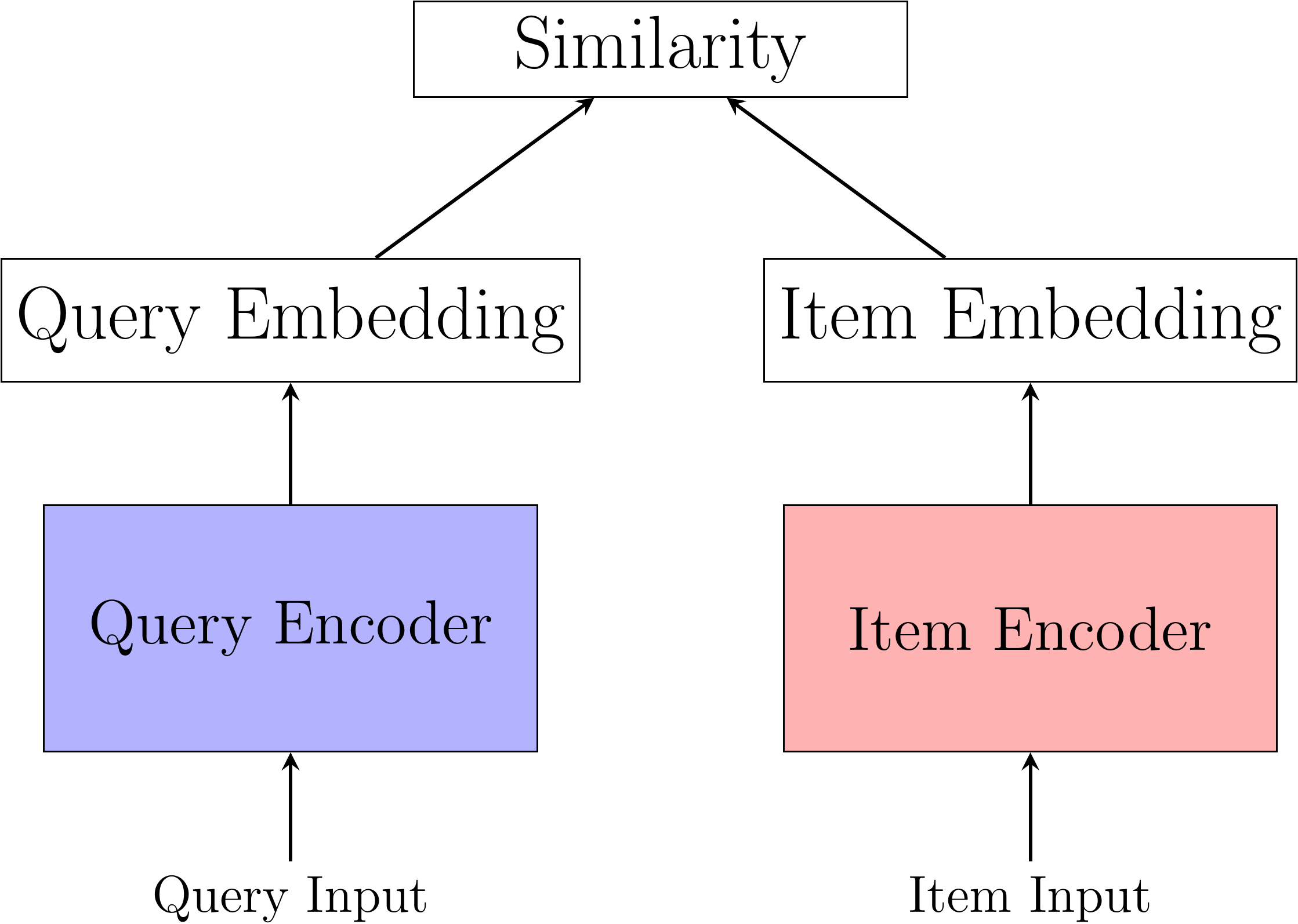}
  \caption{Query-to-item framework: Query embeddings and their similarities are precomputed. Users are represented by a dynamic set of queries that can be updated as needed.}
  \label{fig:item-to-item-framework}
\end{figure}

Our approach overcomes these drawbacks by modelling item-to-item relationships, as illustrated in figure \ref{fig:item-to-item-framework}. Here, the first item represents the query context (an item that has been bought or viewed), while the second item is the target (the item that is subsequently bought). We utilize dual encoders to effectively capture relationships between viewed and bought items, as well as between items bought together. Specifically, our contributions include:

\begin{enumerate}

  \item We demonstrate how a transformer-based two-tower architecture, also known as dual encoders, can be utilized to generate multiple embeddings per item in one model run. Generating multiple embeddings is effective for capturing the various roles of items, and generating them with one model run provides inference efficiency.

  \item We show how we represent customers with multiple queries, where each query corresponds to a product that the customer has interacted with, either through a view or a buy. This approach of representing customers by a set of queries allows us to precompute query embeddings and their respective similarities, facilitating the generation of personalized feeds in near real-time (updates occurring every 2 minutes). This approach offers the benefits of efficiency, as queries are shared, and interpretability, as each recommendation is associated with a specific query.

  \item We showcase real-world applications of our approach in deployed systems at Bol, namely, \textit{Top deals for you}, \textit{Top picks for you}, and \textit{New for you}. By indexing products that are on sale, new or popular and matching them with selected customer query representations, we generate the \textit{Top deals for you}, \textit{New for you}, and \textit{Top picks for you} recommendations.

\end{enumerate}

\section{Related Work} 

Pre-deep learning era, matrix factorization methods were used for personalized recommendations (see \cite{hu2008collaborative, koren2009matrix, su2009survey, koren2022advances}). Since the AlexNet paper \cite{alexnet_NIPS2012_c399862d}, which showed the value of deep learning in image recognition, deep learning has also been applied in recommender systems \cite{he2017neural, zhang2019deep}. Among this rich literature (see survey \cite{zhang2019deep}), the papers most related to our work come from industrial recommender systems such as those of eBay \cite{wang2021personalized}, Youtube \cite{covington2016deep, 10.1145/3298689.3346996}, Google Play \cite{10.1145/3366424.3386195}, Pinterest\cite{10.1145/3534678.3539156, 10.1145/3394486.3403280}, and Alibaba \cite{10.1145/3357384.3357814, 10.1145/3394486.3403344,10.1145/3219819.3219869}. We examine these papers on how they address the customer representation, item representation, and retrieval challenges.

\subsection{Customer Representation Challenge}
The YouTube paper \cite{covington2016deep} uses a Multilayer Perceptron (MLP) model to encode both user and video entities into the same space. The user encoding model takes as inputs embedded video watches (50 recent watches), embedded search tokens (50 recent searches) and user attributes such as age and gender. A vocabulary of 1M videos and 1M search tokens is embedded with 256 floats.

The eBay paper \cite{wang2021personalized} uses a recurrent (GRU) model to generate user embeddings. The inputs to the GRU model are item or query embeddings along with their respective event type embeddings. The event type embeddings are defined by four dimensions and serve to capture various actions on the items. The item embeddings are based on content-based features such as item titles, categories (e.g., mobile phones), and structured aspects (e.g., brand: Apple, network: Verizon). The user embedding has 64 dimensions.

The Pinterest paper \cite{10.1145/3534678.3539156} uses a transformer model to represent the user in 256 dimensions. The inputs to the model are: a sequence of Pins, represented by their PinSage embedding (256-dimensional)  and metadata features: action type, surface, timestamp, and action duration \cite{10.1145/3394486.3403280}. 

To capture the diverse and multifaceted interests of users, prior work from Pinterest and JD.com used multiple embeddings per user \cite{10.1145/3534678.3539156, 10.1145/3394486.3403280, 10.1145/3357384.3357814}. While the notion of employing multiple embeddings to represent users is similar to our method, it also differs. In our solution, the embeddings that constitute customer representations are not unique to each individual customer but rather, are shared among users.

\subsection{Item Representation Challenge}
The YouTube paper \cite{covington2016deep} represents videos with embeddings of 256 dimensions based on Item IDs. The eBay study \cite{wang2021personalized} employs a 3-layer MLP to create item embeddings with a 64-dimensional output. These embeddings are derived from inputs that include title, aspect, and category embeddings. Each of these embeddings is formulated as a Continuous-Bag-of-Words (CBOW) representation, corresponding to the tokens found in the title, aspect, and category. The Pinterest paper \cite{10.1145/3534678.3539156} uses an MLP model to represent items (more specifically, Pins) based only on PinSage embeddings of dimension 256. 

Our work utilizes textual metadata (such as the title and category of a product) to embed item entities. In the YouTube paper, item IDs are used as input to the neural network model, leading to a larger model size due to the need to store an embedding table of significant size. In contrast, our approach generates embeddings directly from input metadata, eliminating the need for a separate table. This is similar to the eBay paper, which also utilizes metadata alone to represent items \cite{wang2021personalized}.

\subsection{Candidate Retrieval Challenge}
\subsubsection{Training challenge}
The most common training strategy for learning user and item embeddings is based on a two-tower user-item framework (see papers from eBay, YouTube and Pinterest \cite{wang2021personalized,covington2016deep,10.1145/3298689.3346996,10.1145/3534678.3539156}). 
The user-item framework tackles the twin challenges of user representation and training using two neural networks in one go. The first network represents user activity of item views and searches whereas the second network represents target items. Variations exist in both the models employed for user and item representation, as well as in the input types fed into the model. Additionally, variations arise in the negative sampling approach utilized during training.

Our training strategy also builds upon the two-tower model and  negative sampling techniques. However, it emphasizes capturing item-to-item relationships, rather than the more common user-to-item relationships. During training for the retrieval stage, our work eliminates the necessity for user specific data and modeling the user, focusing solely on aggregated item-to-item relationships, specifically view-buy or buy-buy interactions.

\subsubsection{Inference challenge}
The approach to overcoming the inference challenge is essentially the same for all large-scale recommender systems. Embeddings of items are indexed and approximate nearest neighbor search is used to efficiently retrieve the most relevant items for given queries represented by user embeddings. Most systems differ in the tools used, e.g., the vector database. For example, eBay uses  FAISS \cite{wang2021personalized}, an open source library from Facebook. Youtube and Pinterest use their own implementations \cite{covington2016deep, 10.1145/3534678.3539156,scam_paper}. Our work uses the FAISS library \cite{johnson2019billion} for indexing and search operations. Since all potential query embeddings (item views and buys) are known in advance, we precompute their similarities and store the query results in a lookup table. Personalized recommendations are then generated by identifying relevant queries for a user and retrieving the corresponding recommendations.

\section{Methodology}

\begin{figure*}[h]
        \centering
        \begin{subfigure}[b]{0.4\textwidth}
            \centering
            \includegraphics[width=\textwidth]{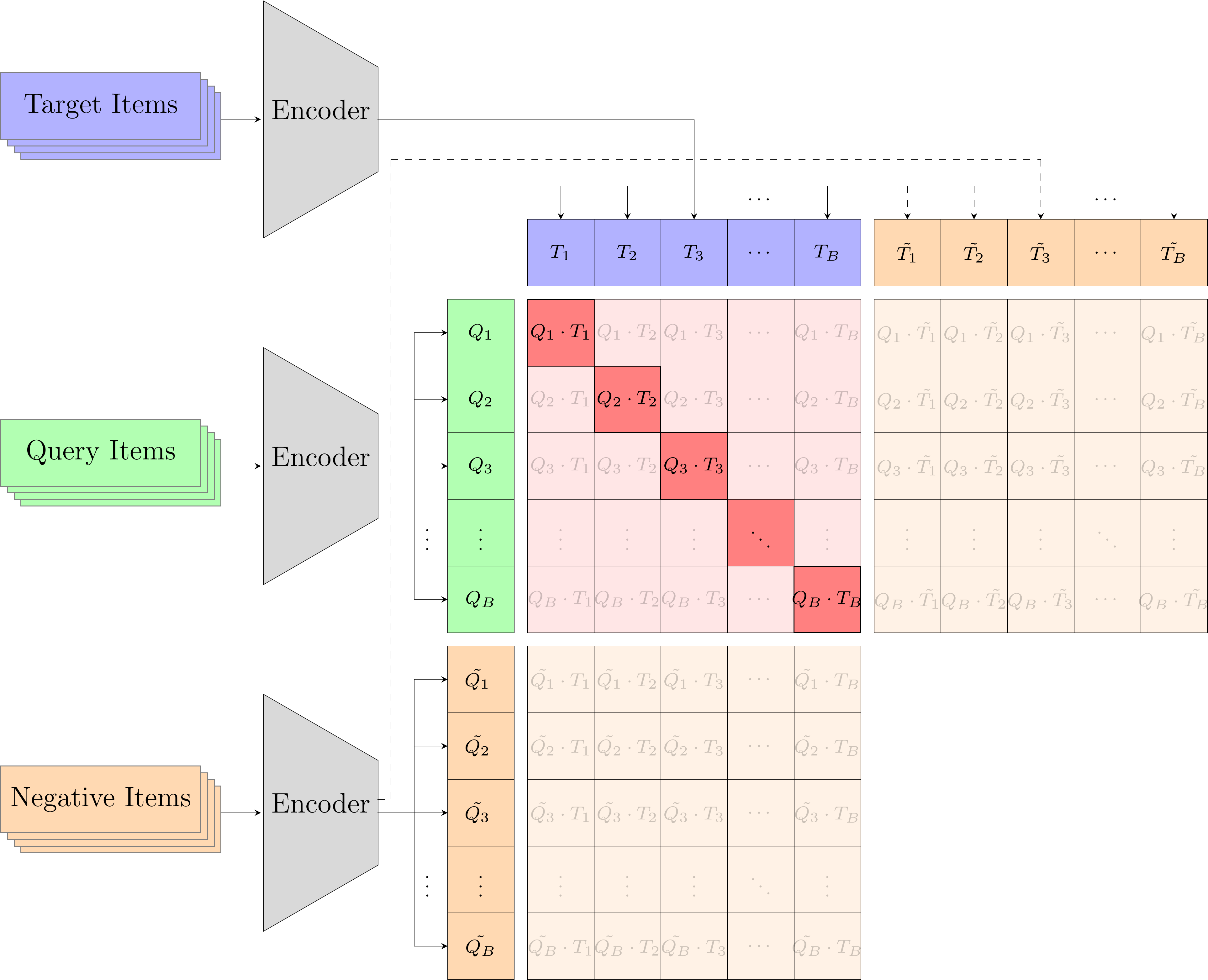}
            \caption[what is this]%
            {{\small Step 1: Contrastive pre-training}}    
            \label{fig:contrastive_pretraining}
        \end{subfigure}
        \hfill
        \begin{subfigure}[b]{0.55\textwidth}  
            \includegraphics[width=0.7\textwidth]{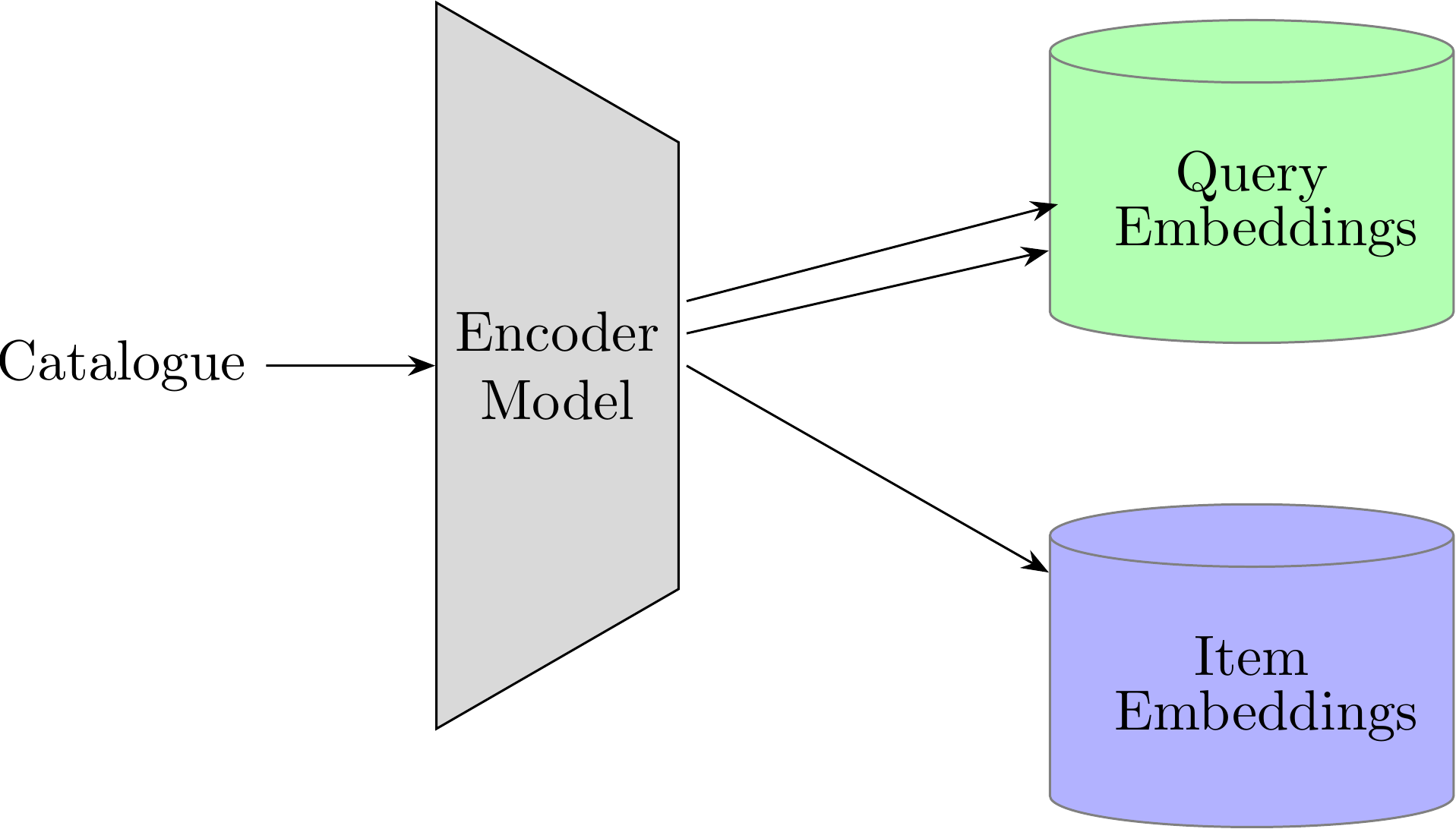}
            \caption[]%
            {{\small Step 2: Generating embeddings}}    
            \label{fig:generating_embeddings}
        \end{subfigure}
        \vskip\baselineskip
        \begin{subfigure}[b]{0.4\textwidth}   
            \centering 
            \includegraphics[width=\textwidth]{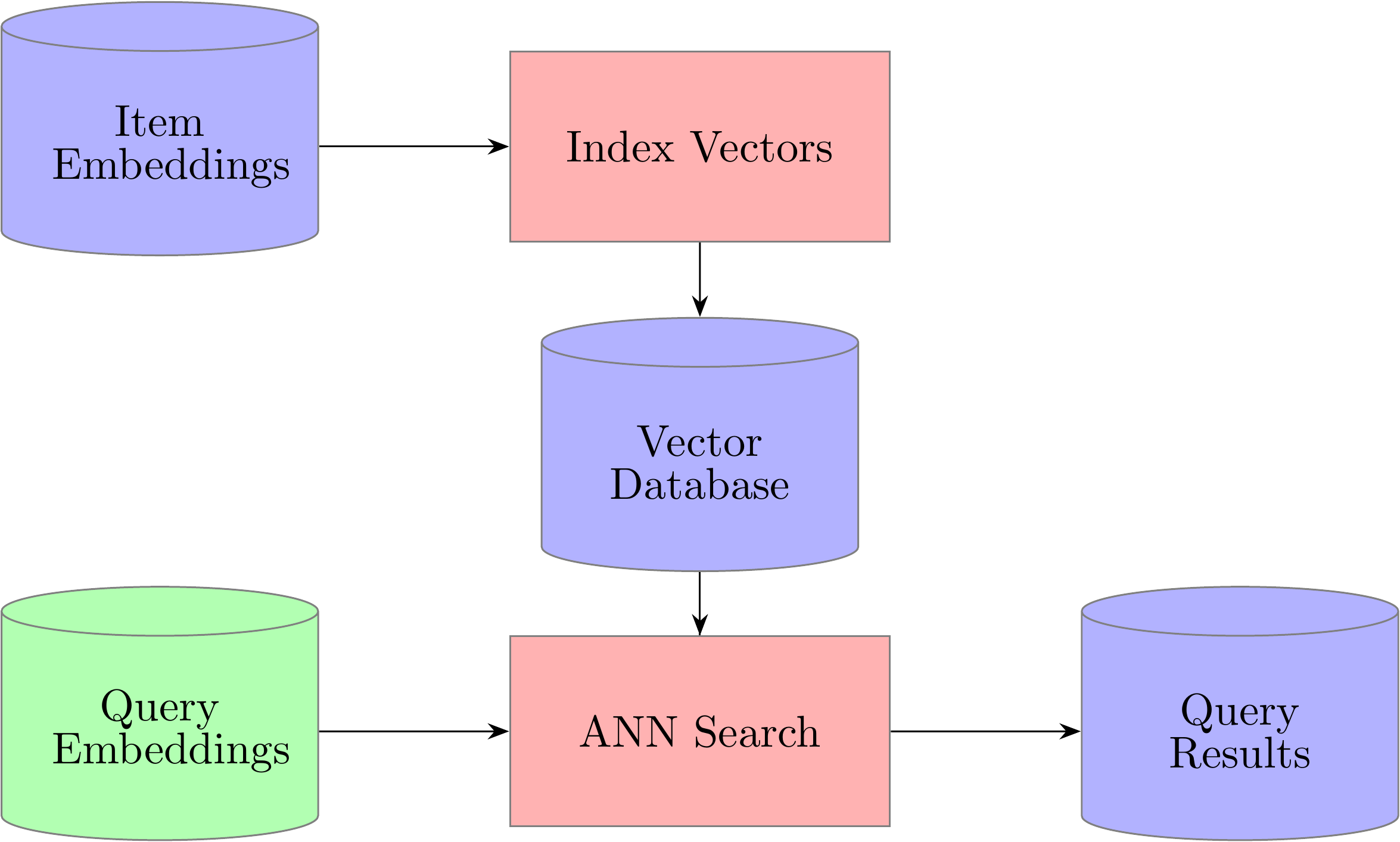}
            \caption[]%
            {{\small Step 3: Indexing and precomputing similarities}}    
            \label{fig:indexing_and_similiarities}
        \end{subfigure}
        \hfill
        \begin{subfigure}[b]{0.55\textwidth}   
            \centering 
            \includegraphics[width=\textwidth]{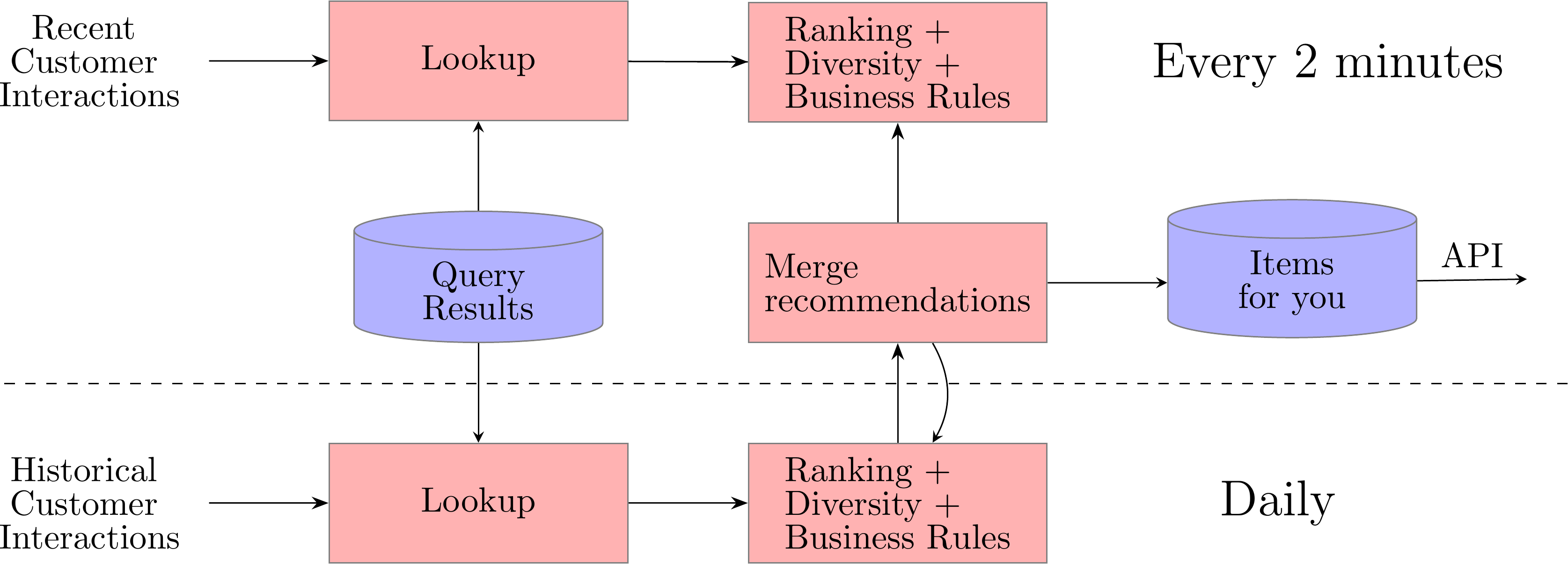}
            \caption[]%
            {{\small Step 4: Generating personalized feed recommendations}}    
            \label{fig:generating_personal_recommendations}
        \end{subfigure}
        \caption[]
        {\small The major steps involved in generating near real-time personalized recommendations} 
        \label{fig:pfeed_workflow_steps}
        
    \end{figure*}

Our method for creating personalized feed recommendations, which we call Pfeed, involves two phases. In the first phase, we train and produce multi-vector item embeddings (see figures \ref{fig:contrastive_pretraining} and \ref{fig:generating_embeddings}). In the second phase, these embeddings are applied to generate personalized product recommendations (see figures \ref{fig:indexing_and_similiarities} and \ref{fig:generating_personal_recommendations}). The goal of the first phase is to capture item-to-item relationships through embeddings. We use "query-to-item" and "query-to-target" interchangeably to refer to the same concept of item-to-item relationships.

\subsection{Representing an Item with Three Embeddings}

In Pfeed, an item can play one of three roles: 1) view query, 2) buy query, and 3) target item. View queries are items clicked during a session leading to the purchase of specific items, thus creating view-buy relationships. Buy queries, on the other hand, are items frequently purchased in conjunction with or shortly before other items, establishing buy-buy relationships. The items that come after view or buy queries are the target items. Our goal is to capture the three roles of an item - view query, buy query, and target - using three distinct embeddings, all generated by a single encoder. 

\subsection{Model Architecture - Generating Three Item Embeddings with One Model Run}
\label{subsec:model_architecture}
\begin{figure}[h]
  \centering
  \includegraphics[width=0.8\linewidth]{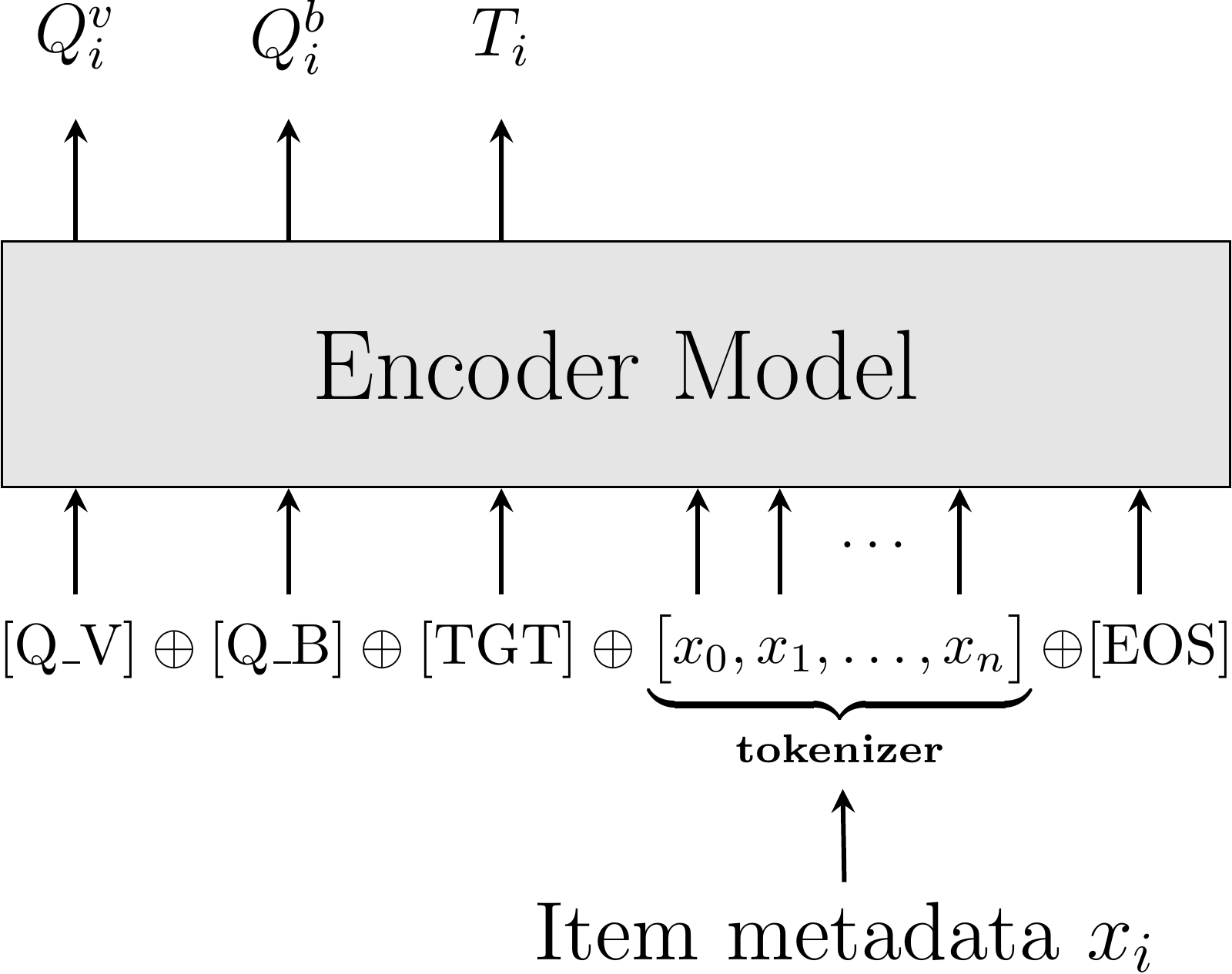}
  \caption{The SIMO (Single Input Multi Ouput) embedding model generates three embeddings per item in one model run using three special tokens: \texttt{[Q\_V]}, \texttt{[Q\_B]}, and \texttt{[TGT]}.}
  \label{fig:inputs_outputs_encoder}
\end{figure}

We use a transformer encoder \cite{NIPS2017_3f5ee243} to generate three embeddings for a given item, each corresponding to the view, buy, or target role. To achieve this, we first tokenize the item metadata into a sequence of tokens using the sentencepiece library \cite{kudo-richardson-2018-sentencepiece}. We then prepend three special tokens: \texttt{[Q\_V]}, \texttt{[Q\_B]} and \texttt{[TGT]} as shown in figure \ref{fig:inputs_outputs_encoder}. These special tokens play a similar role as the \texttt{[CLS]} special token in BERT \cite{devlin2018bert}. 
The first three embeddings from the transformer's final layer, corresponding to the special tokens \texttt{[Q\_V]}, \texttt{[Q\_B]}, and \texttt{[TGT]}, respectively represent the item's view query, buy query, and target embeddings. Because all these three embeddings are generated in one model run, we call the model a Single Input Multi Output (SIMO) embedding model. The SIMO model achieves threefold efficiency compared to a SISO (Single Input and Single Output) embedding model, which requires executing the model three times with distinct prompts for each of the three item roles (view query, buy query, and target roles).

\subsection{Training with Contrastive Learning}
\label{subsec:training_with_contrastive_learning}
\subsubsection{Training data}

We train the SIMO embedding model with query-target pairs consisting of the two types of relationships. The first set consists of item pairs of view-buy relationship (i.e., \{$q$, view, $t$\}). The second set consists of items pairs of buy-buy relationship (i.e., \{$q$, buy, $t$\}). We combine the sets to form one set $\{(q_i,r_{i},t_i)\}^N_{i=1}$, where $(q_i,r_i,t_i)$ corresponds to a positive example, indicating that item $q_i$ and interaction (or relation) $r_i$ led to the purchase of item $t_i$. In addition to query-target item pairs, we also sample random items to reduce bias \cite{yang2020mixed}. 


\subsubsection{Dual encoders}

\begin{figure}[h]
  \centering
  \includegraphics[width=\linewidth]{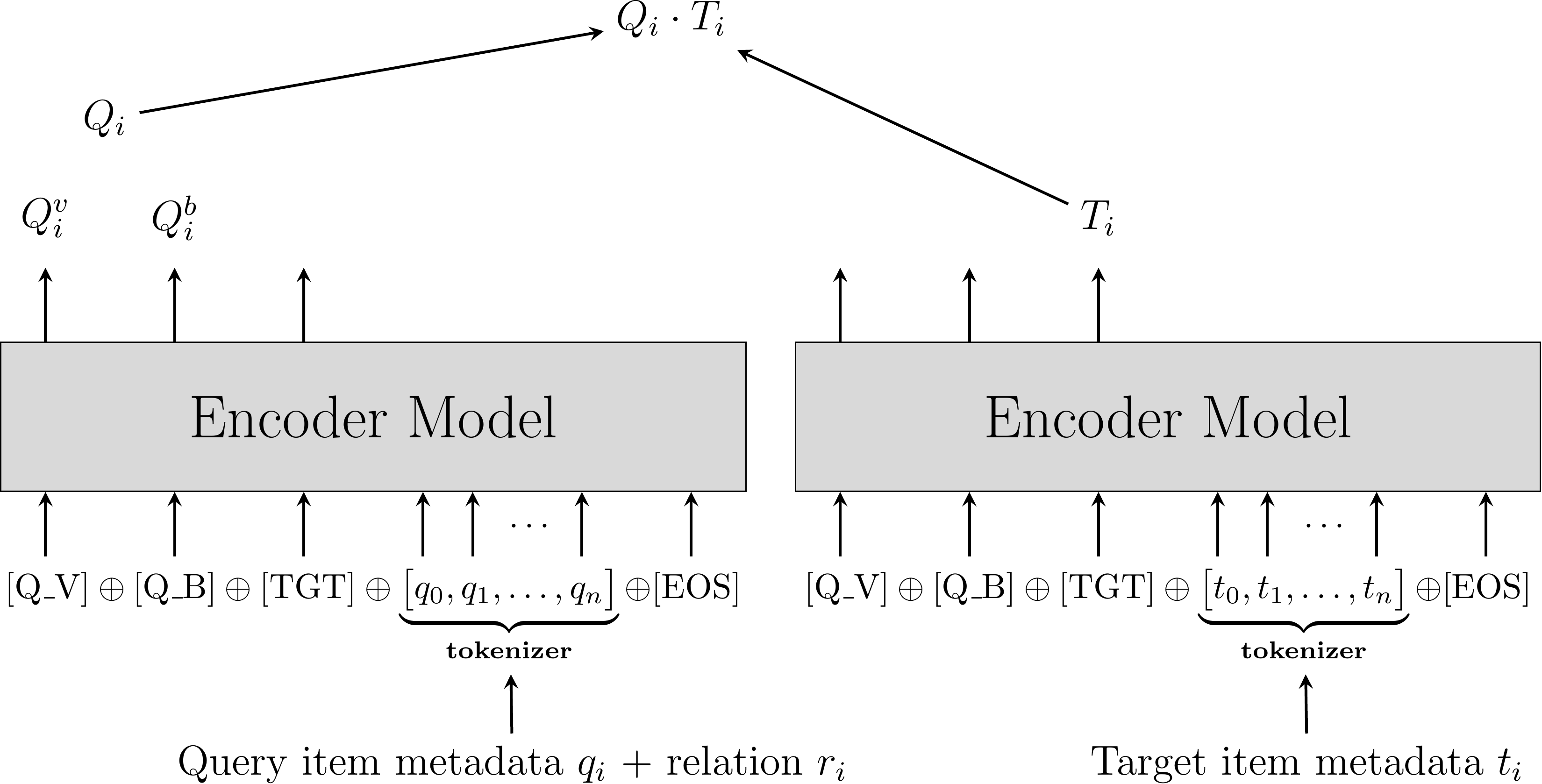}
  \caption{Inputs to dual SIMO encoders: the query encoder takes in the metadata of the query item and generates three embeddings and the target encoder takes in the metadata of the target item and generates three embeddings. During training, the loss is determined by the target embedding derived from the target item $t_i$ encoder and pairing it with a query embedding from the query item $q_i$ encoder, selected by the relation $r_i$ indicator. }
  \label{fig:special_symbols_tokens_dual_encoders}
\end{figure}

The objective of our training is to get a model that produces similar embeddings for matching query-target $(q_i,r_{i},t_i)$ inputs and dissimilar embeddings for non-matching inputs such as $(\tilde{q_i},r_{i},t_i)$ or $(q_i,r_{i},\tilde{t_i})$. To achieve this objective, we employ dual encoders. We feed the query input $q_{i}$ and the target input $t_{i}$ into two instances of the transformer encoder $E$. The encoder $E$ maps $q_{i}$ and $t_{i}$ independently and outputs three embeddings of $Q_i^v$, $Q_i^b$, and $T_i$. From target encoder, we take $T_i$ embedding and do a dot product with the $Q_i$ embedding of the query encoder, which is $Q_i^v$ or $Q_i^b$, depending on the relation $r_i$ (see figure \ref{fig:special_symbols_tokens_dual_encoders}).  When the training samples also include randomly sampled items, called random negatives, we use the same encoder $E$ to generate embeddings: $\tilde{Q}_i^v$, $\tilde{Q}_i^b$, and $\tilde{T}_i$ . These embeddings are mixed with the embeddings of in-batch negatives during training \cite{yang2020mixed}.

\subsubsection{Training objectives}
The training objective consists of two contrastive loss terms. The first loss term employs a query-target softmax formulation (see equation \ref{eq:query_to_target}). In this formulation, we sample negative targets for a given query-target pair. The second loss term employs a target-query softmax (see equation \ref{eq:target_to_query}), where negative queries are sampled for the same query-target pair. We use four types of negative sampling strategies: 1) in-batch negatives, 2) uniformly sampled negatives, and 3) mixed negatives \cite{yang2020mixed} which is a combination of in-batch negatives and uniformly sampled negatives, and 4) self-negatives.

\begin{align}
\mathcal{L}_1 = -\frac 1 {|\mathcal{B}|} \sum_{i=1}^{|\mathcal{B}|} 
\underbrace{\log \frac {e^{\beta \mathbf{Q}_i \cdot \mathbf{T}_i}} { 
\sum_{j=1}^{|\mathcal{B}|} e^{\beta \mathbf{Q}_i \cdot \mathbf{T}_j}
+ \sum_{j=1}^{|\mathcal{N}|} e^{\beta \mathbf{Q}_i \cdot \tilde{\mathbf{T}}_j}
}}_\text{query $\rightarrow$ target softmax}
)
\label{eq:query_to_target}
\end{align}

\begin{align}
\mathcal{L}_2 = - \frac 1 {|\mathcal{B}|} \sum_{i=1}^{|\mathcal{B}|} \underbrace{\log \frac {e^{\beta \mathbf{Q}_i \cdot \mathbf{T}_i}} {
\sum_{j=1}^{|\mathcal{B}|} e^{\beta \mathbf{Q}_j \cdot \mathbf{T}_i} 
+ \sum_{j=1}^{|\mathcal{N}|} e^{\beta \tilde{\mathbf{Q}}_j \cdot \mathbf{T}_i}}}_\text{target $\rightarrow$ query softmax})
\label{eq:target_to_query}
\end{align}

In equations \ref{eq:query_to_target} and \ref{eq:target_to_query}, 
$\mathcal{B}$ represents a batch of embedding pairs for positive samples:
$ \{ (\mathbf{Q}_1, \mathbf{T}_1), (\mathbf{Q}_2, \mathbf{T}_2), \ldots, (\mathbf{Q}_{|\mathcal{B}|}, \mathbf{T}_{|\mathcal{B}|}) \} $. 
$ \mathcal{N} $ represents a set of embeddings from negative items that are uniformly sampled from the catalog and appear as $\tilde{\mathbf{T}}_j$ or $ \tilde{\mathbf{Q}}_j $, depending on the direction of the softmax computation (query-to-target or target-to-query). Each embedding is L2 normalized (i.e., $\|\mathbf{Q}\|_2 = 1$ and $\|\mathbf{T}\|_2 = 1$). The scale parameter $ \beta $ is a parameter that is trained with the model parameters. Initially, we tried a few manually fixed values (e.g., 10, 100) and found it to affect performance significantly. 

\subsection{Inference}
Pfeed has three inference steps: precomputing embeddings, precomputing similarities and generating personalized feeds.
\subsubsection{Precomputing embeddings}
After successful training using the approach described above, we use the resulting trained encoder to generate embeddings for all items in the catalog (see figure \ref{fig:generating_embeddings}). For each item, we generate three embeddings. The first two embeddings are query embeddings for when the item is viewed (indicated as embedding $ Q_i^v $ in figure \ref{fig:inputs_outputs_encoder}) or bought (indicated as embedding $Q_i^b$ in figure \ref{fig:inputs_outputs_encoder}). The third embedding is for when the item is used as a target item (indicated as embedding $T_i$ in figure \ref{fig:inputs_outputs_encoder}).

\subsubsection{Precomputing similarities}
The target embeddings of all items in the catalog (or selected part of it) are indexed with a vector indexing library (in our case, we use FAISS) and we search against the index using the view query and buy query embeddings of all items in the catalog. If the catalog has $ N $ items, then we get $2 \times N$ queries (view and buy for every item in the catalog). For each of the $2 \times N$ queries, we get the $M$ most similar items, resulting in a table with $2 \times N \times M$ entries (see figure \ref{fig:indexing_and_similiarities}). Only entries with a score greater than a prefixed threshold are stored in a lookup table. We fix this threshold from known item-to-item scores (validation data split). Similarity scores above the first percentile (approximately 15\%  of the original set) are stored in the lookup database. 

\subsubsection{Generating personalized feeds}
The process for generating a ranked list of items per customer includes: 1) selecting  queries for each customer (up to 100), 2) retrieving up to 10 potential next items-to-buy for each query, and 3) combining these items and applying ranking, diversity, and business criteria (see figure \ref{fig:generating_personal_recommendations}). This process is executed daily for all customers and every two minutes for those active in the last two minutes. Recommendations resulting from recent queries are prioritized over those from historical ones.

\subsection{Case Study: Personalized Item Feeds at Bol}
We applied Pfeed to generate multiple personalized feeds at Bol, one of the largest e-commerce platforms of the Netherlands and Belgium. The feeds can be seen on the app or website and have titles such as \textit{Top deals for you}, \textit{Top picks for you}, and \textit{New for you}. These feeds differ on at least one of two factors: the specific items targeted for personalization and/or the particular queries selected to represent customer interests.

\subsubsection{Top deals for you}

This feed personalizes items with promotional offers or discounted prices. Pfeed takes the most recent 100 unique customer item views/buys (per category) as query keys. And for each key, it retrieves up to 10 potential discounted items for the customer to buy. This is achieved by accessing precomputed query results and merging them, ensuring near real-time response in the process. This is done daily for all customers and every 2 minutes for recently active customers 
(see figure \ref{fig:generating_personal_recommendations}).

\subsubsection{New for you}

This feed personalizes newly released items. New items, often marked by limited interactions, present a challenge to recommender systems reliant on item IDs or interaction data. However, Pfeed circumvents this cold-start issue because it generates item embeddings using textual catalog metadata \cite{text_is_all_you_need_kdd_2023}. The \textit{New for you} feed works similarly to the \textit{Top deals for you} feed, with the distinction being the type of items selected for personalization. In \textit{New for you}, 
items are designated as new if they fall within the most recent 10\% of items based on their release date, relative to their specific categories. This approach guarantees that each category features its own set of new items, accommodating the varying time scales across different categories.

\subsubsection{X for you}
In general, Pfeed generates \textit{X for you} by limiting the search index or the search output to consist of only items of $X$. In addition to \textit{Top deals for you} and \textit{New for you}, Pfeed has been used to generate other feeds, namely \textit{Top picks for you} and \textit{Select deals for you}. Items for \textit{Top picks for you} come from those that have a certain level of popularity and match the customers' most recent queries from their most frequently interacted with categories. Items for \textit{Select deals for you} come from items that are curated to reward customer loyalty and apply only to customers who are Select members.

\section{Experiments}
To evaluate Pfeed, we run both offline and online experiments. 
The offline experiments are used to evaluate the quality of the embeddings and to illustrate the effects of different design choices on performance. To understand the impact of the embeddings on the personalized feed system, we report results from an online A/B testing experiment. The experiments are specifically designed to answer the following questions.

\begin{description}
\item [Q1:] How does the model that produces three embeddings in one run (SIMO model) compare in terms of performance to the model that generates each embedding in three separate runs (SISO model)?

\item [Q2:] How effective is the SIMO model for cold-start product recommendation? And popular items?

\item [Q3:] How sensitive is the SIMO model to the training strategy, particularly concerning negative sampling and model sizes.

\item [Q4:] How effective are these query-target relationships in generating personalized feeds (online A/B testing)?

\end{description}

\subsection{Dataset}

We create view-buy and buy-buy datasets, comprising of approximately two million positive training/testing samples from around a million unique items (see table \ref{tab:dataset_statistics}). These datasets are constructed from customer item views and item buys. 

\begin{table}[H]
  \caption{Bol dataset statistics}
  \label{tab:dataset_statistics}
  \begin{tabular}{ccc}
    \toprule
     Dataset & \# of positive pairs  & \# of distinct items \\ 
    \midrule
    view-buy & 0.99M  & 1.08M \\
    buy-buy & 0.96M & 0.27M  \\
    Negative & - & 2.00M  \\
    Combined & 1.95M & 3.28M  \\
  \bottomrule
\end{tabular}
\end{table}

\subsubsection{view-buy dataset} 
The view-buy dataset consists of item pairs with view-buy relationships. The pairs are constructed from converting customer sessions. Items that are purchased become target items and the items that were viewed in the same session become the view queries. Of all the view-buy pairs aggregated from sessions from the last four months, we choose the top one million pairs that meet a minimum occurrence threshold and have a high association strength as measured by a cosine metric \cite{10.1145/2723372.2742785}).

\subsubsection{buy-buy dataset} 
The buy-buy dataset consists of item pairs with buy-buy relationships. The pairs are constructed from customer purchases. Items that are purchased later in time become target items and the items that were purchased earlier in time become the buy queries. From all the possible buy-buy pairs constructed from the customer purchases, we select the top one million pairs that meet a minimum occurrence threshold and have a high association strength as measured by a cosine metric.

\subsubsection{Negative dataset} 
In addition to view-buy or buy-buy datasets, we also use a negative dataset that consists of uniformly sampled random items (about two millions). The purpose of this dataset is to reduce selection bias \cite{yang2020mixed}.

\subsection{Offline Evaluation} 
We use the recall metric to compare different design choices. Our dataset is split into training, validation and test sets in the proportions of 80\%, 10\%, and 10\%. To the target items $ t_{i} $ in the test samples $ (q_i, r_i, t_i) $, we add a distractor set $ \tilde C $ of one million items, randomly sampled from the item catalog (a similar approach is used in ItemSage from Pinterest \cite{10.1145/3534678.3539170}). We consider a design choice to be better when its recall@K is higher, i.e.,  the proportion of $ (q_i, r_i, t_i) $ samples for which the retrieved item $ t_i $ is ranked within the top K among $ \tilde C \cup t_i $.

\subsection{Model Architecture Details} 

We use a transformer encoder model with four layers and eight attention heads. The model is identified as SIMO-128, where 128 represents the size of the hidden dimension. Depending on the input sequence we feed to the model, we have either a SIMO or a SISO embedding model.

\subsection{Model Training Details} 
We use Pytorch and Pytorch lightning for training the transformer model. The model is optimized with Lamb optimizer \cite{you2019large} with a learning rate of 0.01 on four V100 GPUs using Distributed Data Parallel (DDP) strategy. Each GPU runs three instances of the model, each handling a batch size of 1024. These instances handle input sequences from query, target, and negative item sequences after tokenization using the sentencepiece library \cite{kudo-richardson-2018-sentencepiece} using a vocabulary size of 20k. Prior to loss computation, all forward passes from each GPU are gathered, resulting in a total batch size of $1024 \times 4 (=4096)$. The loss is computed by incorporating both in-batch and uniformly sampled negative samples, amounting to a total of $8192$ minus $1$ negatives per positive sample \cite{yang2020mixed}. To stabilize training, gradients are clipped to 0.5. The context length of the input sequence is fixed to a maximum of 64 tokens, sufficient for encoding item titles and essential metadata such as categories but excluding descriptions.
 
\subsection{Retrieval Performance and Efficiency (Q1)}
The query-target retrieval system, based on the embeddings generated by a transformer model that generates three embeddings with a single run (SIMO embedding model), performs comparably to the model that generates the embeddings separately (SISO embedding model). The SIMO embedding model generates embeddings three times faster than the SISO embedding model (see table \ref{tab:recall_on_test_datasets}). 

\begin{table}[H]
  \caption{Recall@K on view-buy and buy-buy datasets. The SIMO-128 model performs comparably to the SISO-128 model while being 3 times more efficient during inference.}
  \label{tab:recall_on_test_datasets}
  \begin{tabular}{cccc}
    \toprule
      \multirow{ 2}{*}{Model} & \multicolumn{2}{c}{Recall@10 (\%)} \\
      \cmidrule(rl){2-3} \\ 
      & view-buy dataset & buy-buy dataset & efficiency \\ 
    \midrule
    SIMO-128  & 41.86 & 36.41 & 3x \\
    SISO-128 & 41.57 & 36.12  & x\\
  \bottomrule
\end{tabular}
\end{table}

\subsection{Retrieval Performance on Cold-start and Popular Items (Q2)}

The query-target retrieval system, based on the SIMO-128 model, shows varying performance depending on the nature of the dataset and the level of popularity of the items. On the buy-buy dataset, recall scores are lower for head items. On the view-buy dataset, recall scores are slightly higher for head items (see table \ref{tab:recall_and_popularity}). This recall score difference between the two datasets is attributed to the differing distributions of query-to-target relationship categories. On the buy-buy dataset, approximately 75\% of the relationships are either one-to-many, many-to-one, or many-to-many (complex relationships). In contrast, on the view-buy dataset, such relationships constitute less than 21\% (see table \ref{tab:relation_type_distributions}). A detailed analysis of recall scores segmented by relationship category reveals a consistent trend across both datasets: scores on item pairs with complex relationships are lower (see table \ref{tab:recall_and_relation_types}). The reasons for this are twofold: First, single vectors face difficulties in capturing complex relationships. Second, during training, the model is inaccurately penalized for failing to replicate the exact query-target pairs provided, rather than being evaluated on its ability to identify any valid query-target pairs.

\begin{table}[H]
  \caption{Impact of item popularity on Recall@K. Performance on popular items is lower than on tail items on the buy-buy dataset. This is due to a higher proportion of complex relations on the buy-buy dataset, indicated in table \ref{tab:relation_type_distributions}. }
  \label{tab:recall_and_popularity}
  \begin{tabular}{ccc}
    \toprule
      \multirow{ 2}{*}{Popularity} & \multicolumn{2}{c}{Recall@10 (\%)} \\
      \cmidrule(rl){2-3}
      & view-buy dataset & buy-buy dataset \\ 
    \midrule
    Cold-start & 38.52 &  59.76  \\
    Tail & 41.66 & 55.88  \\
    Head & 42.32 & 25.54  \\
    All & 41.86 & 36.41  \\
  \bottomrule
\end{tabular}
\end{table}

\begin{table}[H]
  \caption{Relationship categories and their distributions. The buy-buy dataset has a higher distribution of complex relations ($1 \times n$, $m \times 1$ and $m \times n$ relations). }
  \label{tab:relation_type_distributions}
  \begin{tabular}{ccc}
    \toprule
      \multirow{ 2}{*}{Relationship category} & \multicolumn{2}{c}{Distribution (\%)} \\
      \cmidrule(rl){2-3}
      & view-buy dataset & buy-buy dataset \\ 
    \midrule
    $1 \times 1$ & 80.5  & 24.7  \\
    $1 \times n$ & 6.9  & 16.2 \\
    $m \times 1$ & 11.5 & 20.5   \\
    $m \times n$ & 1.1  &  38.6  \\
    All & 100.0 & 100.0  \\
  \bottomrule
\end{tabular}
\end{table}

\begin{table}[H]
  \caption{Relationship categories and Recall@K. Performance is higher on test data with simple $ 1 \times 1 $ relations than with complex relations. The buy-buy dataset has a higher proportion of complex relations ($\sim75\% $), see table \ref{tab:relation_type_distributions}.}
  \label{tab:recall_and_relation_types}
  \begin{tabular}{ccc}
    \toprule
      \multirow{ 2}{*}{Relationship category} & \multicolumn{2}{c}{Recall@10 (\%)} \\
      \cmidrule(rl){2-3}
      & view-buy dataset & buy-buy dataset \\ 
    \midrule
    $1 \times 1$  & 42.08  & 58.01  \\
    $1 \times n$  & 40.22  & 41.98 \\
    $m \times 1$ & 41.71 & 35.55   \\
    $m \times n$  & 37.63 &  20.72  \\
    All & 41.86 & 36.41  \\
  \bottomrule
\end{tabular}
\end{table}

\subsection{Sensitivity of the Retrieval Performance (Q3)}
We conduct a sensitivity analysis of our method by varying the hidden dimensions of the SIMO model and altering particular aspects of the training strategy, particularly the negative sampling strategy. 

\subsubsection{Hidden dimension}
We vary the hidden dimension of the model between 64, 128, 256, 384, and 512 while keeping the rest of the transformer model and training strategy the same. Performance increases as the dimension increases until 384. At dimension 512, the model's performance drops (see table \ref{tab:impact_of_hidden_vector_size}). 

\begin{table}[H]
  \caption{Impact of hidden dimension vector size on Recall@K}
  \label{tab:impact_of_hidden_vector_size}
  \begin{tabular}{cccc}
    \toprule
      \multirow{ 2}{*}{Vector size} & \multirow{ 2}{*}{Parameter \#} & \multicolumn{2}{c}{Recall@10 (\%)} \\
      \cmidrule(rl){3-4}
     &  & view-buy  & buy-buy dataset \\ 
    \midrule
    $64$   & 1.5M & 37.87 & 32.09 \\
    $128$  & 3.6M  & 41.86 & 36.41 \\
    $256$  & 9.1M  & 44.31  & 40.73  \\
    $384$  & 16.6M  & 44.71 & 41.61  \\
    $512$  & 26.0M  & 41.23 & 38.93  \\
  \bottomrule
\end{tabular}
\end{table}

\subsubsection{Negative sampling strategy}
We use four types of negative sampling strategies: in-batch negative sampling, uniform negative sampling, mixed negative sampling, and self-negative sampling. The best performance is achieved with mixed negative sampling, where both in-batch and uniform sampled negatives are used \cite{yang2020mixed}. In-batch negative sampling is second best (see table \ref{tab:impact_of_negative_sampling_strategy}).

\begin{table}[H]
  \caption{Impact of negative sampling strategy on Recall@K} 
  \label{tab:impact_of_negative_sampling_strategy}
  \begin{tabular}{ccc}
    \toprule
      \multirow{ 2}{*}{Negative Sampling} & \multicolumn{2}{c}{Recall@10 (\%)} \\
      \cmidrule(rl){2-3}
      & view-buy dataset & buy-buy dataset \\ 
    \midrule
    Mixed & 41.86 & 36.41 \\
    In-batch  & 40.87  & 35.88  \\
    Uniform   & 39.15 & 31.73 \\
    Mixed + self-negatives & 40.45 & 31.24  \\
  \bottomrule
\end{tabular}
\end{table}
Self-negatives refer to instances where the target embeddings of query items serve as their own negatives (or the query embeddings of target items serve as their own negatives). Self-negatives are advantageous for handling asymmetrical buy-buy relationships or instances of non-repeat purchases. When we add self-negatives from query-item pairs having buy-buy relations to the mixed negatives, we observe a decline in the overall recall score. This suggests that such relationships are less prevalent in the dataset.

\subsection{Online A/B testing (Q4)}

We ran an online A/B testing experiment where we compared a treatment group receiving personalized \textit{Top deals for you} item lists (generated by Pfeed) against a control group that received a non-personalized \textit{Top deals} list, curated by promotion specialists. This experiment was conducted over a two-week period with an even 50-50 split between the two groups. The results showed a statistically significant increase in performance for the treatment group: there was a 4.9\% increase in conversion rates and a 27\% increase in the number of items added to wish lists (see table \ref{tab:ab_test}). Following these results, Pfeed has been deployed and can be found on both the mobile app and the website of Bol.

\begin{table}[H]
  \caption{Online A/B test}
  \label{tab:ab_test}
  \begin{tabular}{ccc}
    \toprule
    Model & Wish list additions & Conversion\\
    \midrule
    Non-personalized deals  & 0.00 & 0.00 \\
    Top deals for you & +27\% & +4.9\% \\
  \bottomrule
\end{tabular}
\end{table}

\section{Conclusions}

In this paper, we introduced Pfeed, a method for generating personalized product feeds on e-commerce platforms. The method has been deployed at Bol with services called: \emph{Top deals for you}, \emph{Top picks for you}, and \emph{New for you} and achieved a significant conversion uplift. Pfeed uses a query-to-item framework as opposed to user-item, the framework most dominant for personalized recommender systems. We highlighted three benefits of the query-to-item framework. 1) Simplification of real-time deployment, as query results can be precomputed and user interests can dynamically be updated in real-time, all without requiring model inference or the unlearning of past preferences. 2) Enhanced interpretability, as each recommendation in the feed can be traced to specific queries. 3) Increased computational efficiency due to the reuse of queries among users. Additionally, we demonstrated the use of multiple special tokens as input in the transformer model, enabling a single model run to generate multiple embeddings for each item. 

\section{Future Work}
Pfeed's embedding approach can be enhanced in two ways: 1) better handling of query-to-item training samples having complex relations and 2) explicit modeling of memorization and generalization features.

\textbf{Modeling complex query-to-item relations}: Pfeed's current method of representing users with a set of individual queries provides flexibility but falls short in modeling sequential user behavior. This isn't inherently an issue, as the ranking phase can incorporate sequential information. However, it requires the embedding-based retrieval phase to be expressive enough to handle an increased set of relevant items, including those that might otherwise be excluded by sequential modeling.
For example, if a user buys diapers, there are numerous potential next purchases such as items related to baby toys or clothes. Pfeed's embedding strategy struggles to model such complex relations (one-to-many, many-to-one and many-to-many relations). In practice, Pfeed settles with the most probable next purchase and thus provides less variety per query. Future enhancements could involve multi-vector query representations, allowing for a wider range of item choices.

\textbf{Modeling memorization and generalization features}: Pfeed's embedding strategy leverages item content, like titles, which is good for generalization, but it does not explicitly incorporate memorization features such as item IDs or popularity. This limitation could impact the system's performance, particularly with popular items. Future work could focus on designing an architecture that can adaptively use memorization features when available, while still relying on generalization features in their absence. This improvement would enable the system to more accurately predict next-item choices, covering both popular and long tail items.

\section{Acknowledgments}

We are grateful to Tim Akkerman, Cuong Dinh, Isaac Sijaranamual, Paulo Barchi, Barrie Kersbergen, Chongze Jiao, Haider Ali Afzal, Bart van de Garde, and Charles de Leau for their suggestions, comments, corrections, and inspiration.




\bibliographystyle{ACM-Reference-Format}
\bibliography{references_acmart}


\end{document}